\documentclass[12pt,preprint]{aastex}

\newcommand{\etal}{{et\,al.}}  \newcommand{\eg}{e.\,g. }

\def\unit #1{\,{\rm #1}} 
\def\kmps{\unit{km \, s^{-1}}}
\def\re{r_{\rm e}}
\def\ibzero{I_{\rm b}(0)}
\def\ibr{I_{\rm b}(r)}
\def\rd{r_{\rm d}}
\def\idr{I_{\rm d}(r)}

\newcommand{\myequn}[1]{\begin{equation}{#1}\end{equation}}

\def\idzero{I_{\rm d}(0)}

\shorttitle{Lenticular Galaxy Formation}
\shortauthors{Barway et al.}

\begin{document}

\title{Lenticular Galaxy Formation - Possible Luminosity Dependence}

\author{Sudhanshu Barway\altaffilmark{1}, 
Ajit Kembhavi\altaffilmark{1}, 
Yogesh Wadadekar\altaffilmark{2},
C.D. Ravikumar\altaffilmark{3},
and 
Y. D. Mayya\altaffilmark{4}
}

\altaffiltext{1}{IUCAA, Post Bag 4, Ganeshkhind, Pune 411007; sudhan@iucaa.ernet.in, akk@iucaa.ernet.in} 
\altaffiltext{2}{Department of Astrophysical Sciences, Princeton University, Princeton, NJ 08544; yogesh@astro.princeton.edu} 
\altaffiltext{3}{Department of Physics, University of Calicut,
Kerala, India, 673635; ravi@iucaa.ernet.in} 
\altaffiltext{4}{Instituto Nacional de Astrofisica, Optica y Electronica,
Luis Enrique Erro 1, Tonantzintla, Apdo Postal 51 y 216, C.P. 72000,
Puebla, M\'exico; ydm@inaoep.mx}

\begin{abstract}
We  investigate the  correlation  between the  bulge effective  radius
($\re$) and disk  scale length ($\rd$), in the  near-infrared $K$ band
for lenticular galaxies in the field and in clusters. We find markedly
different  relations  between the  two  parameters  as  a function  of
luminosity.   Lenticulars with total  absolute magnitude  fainter than
$M_T=-24.5$ show  a positive correlation, in line  with predictions of
secular formation  processes for the  pseudo bulges of  late-type disk
galaxies.    But  brighter  lenticulars   with  $M_T<-24.5$   show  an
anti-correlation,  indicating  that they  formed  through a  different
mechanism. The available  data are  insufficient to reliably determine 
the effect of galaxy environment on this correlation.
\end{abstract}
  
\keywords{galaxies: photometry  --- galaxies: formation  --- galaxies:
fundamental parameters}

\section{Introduction}

Lenticular (S0) galaxies form a morphological transition class between
ellipticals and early-type spirals in the Hubble (1936) classification
system. They  have disks with  luminosity ranging from ten  to hundred
percent of the bulge. The use of lenticulars as a transition class may
be justified by  the fact that in many  of their observable properties
such  as $B/T$  ratio,  colors and  spectral  properties, neutral  and
molecular gas  fraction, star formation rates,  average luminosity and
$M/L$  ratio,  the  more  luminous lenticulars  (which  dominated  the
magnitude  limited  samples during  Hubble's  time)  straddle the  gap
between ellipticals and spirals. When comparing properties, it is  
found that  the  bulges  of  lenticulars  are  very  similar  to  elliptical
galaxies, while  their disks have  similarities to the disks  of early
type spiral  galaxies, except that  they lack conspicuous  spiral arms
(see the review by Fritze -- v. Alvensleben 2004). It has recently been 
suggested by Bedregal,
Arag{\'o}n-Salamanca,  \& Merrifield (2006)  and Arag{\'o}n-Salamanca,
Bedregal, \&  Merrifield (2006) that  a significant fraction
of lenticulars  are faded spirals. Our understanding  of the formation
and  evolution of  lenticular  galaxies, in  terms  of the  individual
physical processes  involved, is  still unclear, inspite  of extensive
efforts  both by observational  and theoretical  means (van  den Bergh
1976, 1990; Bothun 1982, Bekki \etal \ 2002).

Over the last decade, it  has become increasingly clear that the bulge
component  of  galaxies comes  in  two  varieties,  distinct in  their
properties  and parameter  correlations  (Ravikumar \etal  \ 2006  and
references  therein).   Available  evidence  indicates that  the  {\it
classical bulges}  found in ellipticals  and early type  disk galaxies
formed at  an early cosmic  epoch from the rapid
transformation events caused by hierarchical clustering and merging of
comparable  mass galaxies.  Such  bulges typically  are old  and often
show   higher  anisotropic  velocity   dispersion  in   their  stellar
kinematics.  Their  light profile is  usually well represented  by the
S\'ersic (1968)  law with  larger $n$ values $(n \sim 4)$,  and they
often show boxy isophotes. Classical bulges show correlations in their
global  photometric  and dynamical  properties,  like the  Fundamental
plane  (Djorgovski \&  Davis  1987,  Dressler \etal  \  1987) and  the
Photometric plane (Khosroshahi \etal \ 2000a, Ravikumar \etal \ 2006),
reflecting  their virialized nature  or maximized  entropy (Lima-Neto,
Gerbal, \& M{\'a}rquez \ 1999).

The  younger, bluer  bulges found  in late-type  disk  galaxies, which
usually  have  exponential light  profiles  and  disky isophotes,  are
largely rotationally supported. These  weaker {\it pseudo bulges} (see
detailed  review by  Kormendy \&  Kennicutt 2004)  likely formed  in a
dissipational process  due to vertical dynamical  instabilities of the
disks (Combes  \etal \ 1990; Hasan,  Pfenniger, \& Norman  \ 1993) and
subsequently  grew through secular  evolution by  gas infall  into the
central  regions,  possibly  channeled  through a  bar.   The  secular
evolution  process   may  be  episodic   and  may  still   be  growing
pseudo-bulges in late type  disk galaxies today (Kormendy \& Kennicutt
2004).

Given  that  lenticulars  are  intermediate  between  ellipticals  and
early-type spirals in many of  their properties, one might expect that
the bulge  component of lenticulars  is of the classical  type. N-body
simulations support the view  that the spheroidal component of massive
and more  luminous lenticulars is formed  by the major  merger of disk
systems  in a  prograde or  retrograde  encounter (Bekki  1995), in  a
process  similar   to  the  merger  driven   formation  processes  for
ellipticals and  bulges of early-type spiral  galaxies.  Such mergers,
after a  few Gyr,  show elliptical galaxy  like colors and  spectra in
evolutionary  synthesis models  (Fritze --  v. Alvensleben  \& Gerhard
1994a,b). However, such a merger  scenario is only viable in the field
or  within  infalling groups  where  galaxy  encounter velocities  are
sufficiently low  for efficient  merging. Minor mergers  (or accretion
events), on the other hand, are a viable route to the formation of low
luminosity lenticulars  as demonstrated by  N-body simulations (Barnes
1996; Bekki 1998). In such  encounters, stripping of gas from the halo
and disk of  spiral galaxies (Bekki \etal \ 2002)  is accompanied by a
change in  morphology. For such galaxies, the  standard arrangement of
lenticulars on the Hubble sequence may need to be revised.  A suitable
classification for low luminosity  lenticulars may be the one proposed
by van den  Bergh (1979, 1998) wherein the  lenticular galaxies form a
sequence  parallel  to  spirals  with  S0a, S0b,  and  S0c  being  the
gas-stripped  lenticular   analogues  of  Sa,  Sb,   and  Sc  galaxies
respectively. If  indeed a fraction of lenticulars  formed from spiral
galaxies,  the $B/T$  flux ratio  should be  comparable  between these
analogous  classes. Recently,  Laurikainen \etal  \,(2006)  have found
lenticular  galaxies where  the $B/T$  flux ratio  is as  small  as in
typical Sc-type spirals.

Which  of the  models of  bulge formation  operates as  a  function of
galaxy  luminosity and environment  may be  tested by  confronting the
predictions  of bulge  formation  models and  N-body simulations  with
observations.  Bulge  formation models make  some testable predictions
mostly in the form of  correlations expected in the global photometric
properties of  bulges and disks.  For example, secular  bulge formation
models predict  that the  bulge and disk  scale lengths  be correlated
(Martinet 1995; Combes 2000; see\, Carollo, Ferguson, \& Wyse 1999 for
comprehensive review  articles). On  the other hand,  classical bulges
exhibit correlations such as the fundamental plane and the photometric
plane mentioned  above. Such predictions can  be tested quantitatively
by  measuring these  parameters  using one  of  the several  available
bulge-disk decomposition  codes (\eg Wadadekar,  Robbason, \& Kembhavi
1999; Peng \etal \ 2002).

In  this  Letter,  we  examine correlations  amongst  the  photometric
parameters of bulge and disk components of lenticular galaxies, in the
near-infrared $K$ band, as a probe of likely formation mechanisms. Our
goal is to investigate the  relative importance of mergers and secular
evolution in  the formation of the  bulge component, as  a function of
luminosity and environment. Throughout  this Letter, we use $H_{\rm 0}
= 70\,\kmps\,\unit{Mpc^{-1}}$.

\section{The Data and Decomposition Technique}
Our sample  consists of a set  of 35 bright  field lenticular galaxies
from Barway  \etal \ (2005), observed  in the $K$ band  using the 2.1m
telescope  at  Observatorio Astronomico  Nacional,  San Pedro  Martir,
Mexico.   The  original  sample  contained  40  lenticulars  galaxies,
selected  from  the Uppsala  General  Catalogue  (UGC), with  apparent
magnitude  $B  <  14$,  angular  diameter  $D_{25}  <  3$  arcmin  and
declination $5^{\circ}<\delta  < 64^{\circ}  $. The sample,  while not
complete,  is  representative of  bright  field lenticulars.   Further
details on sample selection, observation and data reduction procedures
can be found in Barway \etal \ (2005).

We extracted  the bulge  and disk parameters  for our  sample galaxies
using  a two-  dimensional bulge-disk  decomposition  technique, which
employs a  $\chi^2$ minimization  algorithm as described  in Wadadekar
\etal \  (1999). We decomposed all  the galaxies in our  sample into a
bulge  component with  surface  brightness distribution  given by  the
r$^{1/n}$ law (S\'ersic 1968) with
\myequn{\ibr = \ibzero e^{-2.303b_n{(r/\re)^{1/n}}},}
where  $n$ is  the  S\'ersic  index, $\ibzero$  is  the bulge  central
intensity and the constant $b_n$ is chosen such that $\re$ is the half
light radius for every value of $n$; $b_n$ is the root of the equation
\myequn{P(2n,2.303b_n) = 0.5, \nonumber}
where  $P$($a$,$x$) is the  incomplete gamma  function (see  \eg Press
\etal \ 1992).

The  disk profile  is  well  approximated by  an  exponential $\idr  =
\idzero  e^{-(r/\rd)}$,  where $\rd$  is  the  disk  scale length  and
$\idzero$  is  the  disk  central  intensity.   Apart  from  the  five
parameters mentioned above,  the fit also involves the  bulge and disk
ellipticities and the sky  background. Intensity models convolved with
the  appropriate  point-spread function  (PSF)  determined from  stars
present in the galaxy image are fitted to the observed images so as to
minimize $\chi^2$.  Of the 40 lenticulars in Barway \etal \ (2005), we
obtained satisfactory fits for the  35 galaxies which we have included
in the present analysis.

We augment  our sample with data  from Bedregal, Arag{\'o}n-Salamanca,
\& Merrifield (2006; hereafter BAM06) on 49 lenticular galaxies. These
are relatively  faint objects  with sufficient rotational support for
the disks.  BAM06  used the Two Micron All  Sky Survey (2MASS; Jarrett
\etal  \  2003) data  in  the  $K_s$ band  to  obtain  bulge and  disk
parameters  using the  decomposition code  by Simard  \etal  \ (2002),
assuming S\'ersic  and exponential laws  for the bulge and  disk light
distributions, as we have done with our sample.

\begin{figure}
\plotone{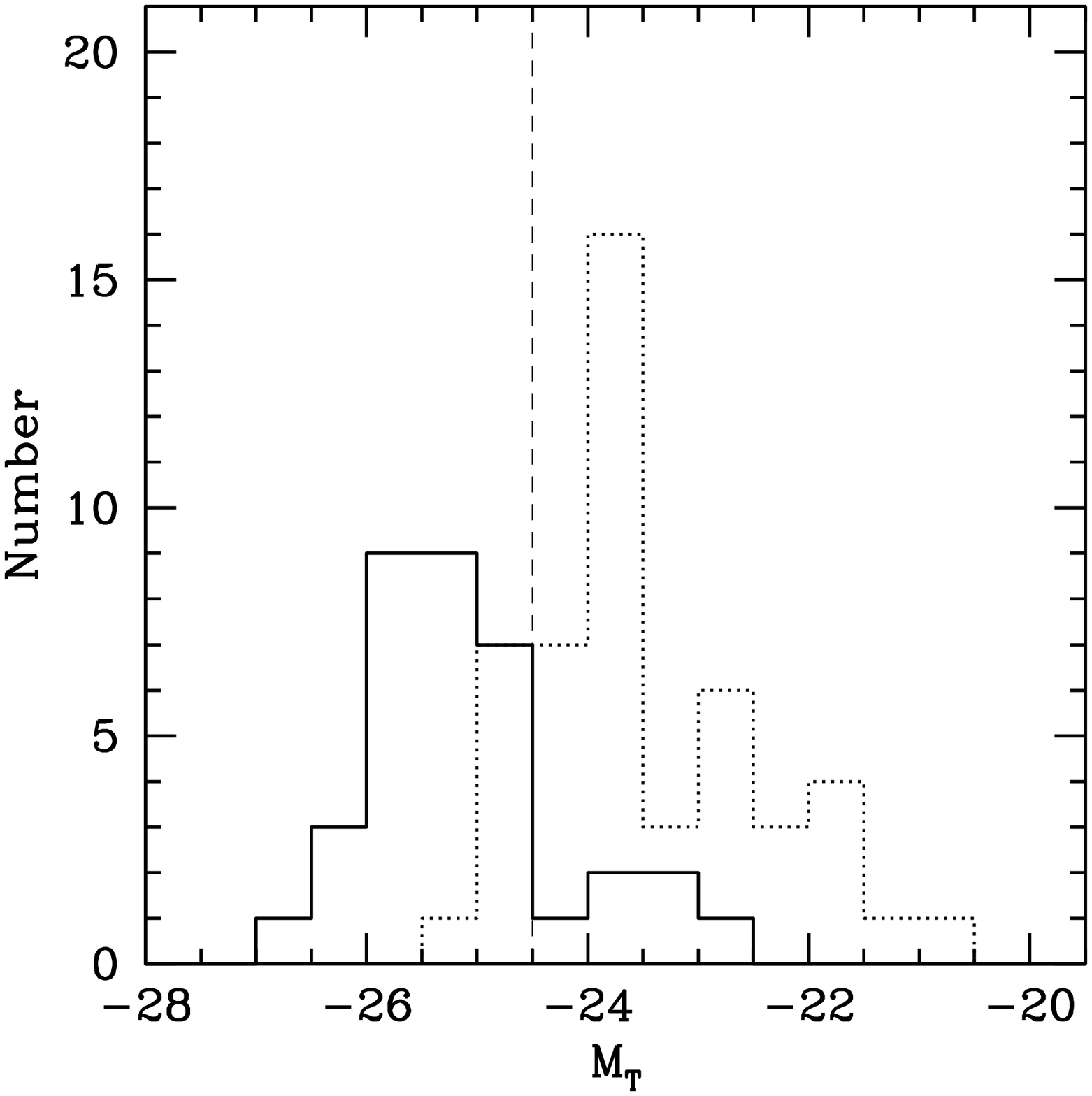}
\caption{Distribution of total absolute magnitude ($M_T$) in $K$ band for our sample (solid line) and for BAM06 lenticulars (dotted line). Vertical dashed line corresponds to total absolute magnitude $ M_T = -24.5$, which we use to separate low and high luminosity lenticulars.}
\label{f1}
\end{figure}

The BAM06 galaxies  complement our sample in two  ways: They extend to
fainter   luminosities   and    provide   lenticulars   in   different
environments. The  sample contains galaxies from the  Coma (14), Virgo
(8),  and Fornax  (6) clusters  along with  21 field  lenticulars.  In
Figure~\ref{f1} we  show the distribution of  total absolute magnitude
($M_T$) in the $K$ band for the combined sample, which is seen to span
a wide range in luminosity.

\section{Bulge-Disk Correlations}
Correlations  between  the  bulge  and  disk  parameters  can  provide
important information  on their interplay and  evolution. Courteau, de
Jong,  \&   Broeils  (1996)  reported  a   correlation  between  bulge
half-light radius (effective radius) $\re$ and disk scale length $\rd$
for late-type spiral galaxies (see also de Jong 1996), which was later
confirmed by  Khosroshahi, Wadadekar, \& Kembhavi  (2000b).  Bulge and
disk  size correlations are  well understood  in models  in which  the disk
forms first and the bulge  emerges from the disk, via angular momentum
redistribution  processes (Combes \etal  \ 1990;  Saio \&  Yoshi 1990;
Struck-Marcell  1991).  On  the  contrary, models  which predict  disk
formation from an existing central bulge, do not support the idea of a
strong bulge-disk size correlation.

\begin{figure}
\plottwo{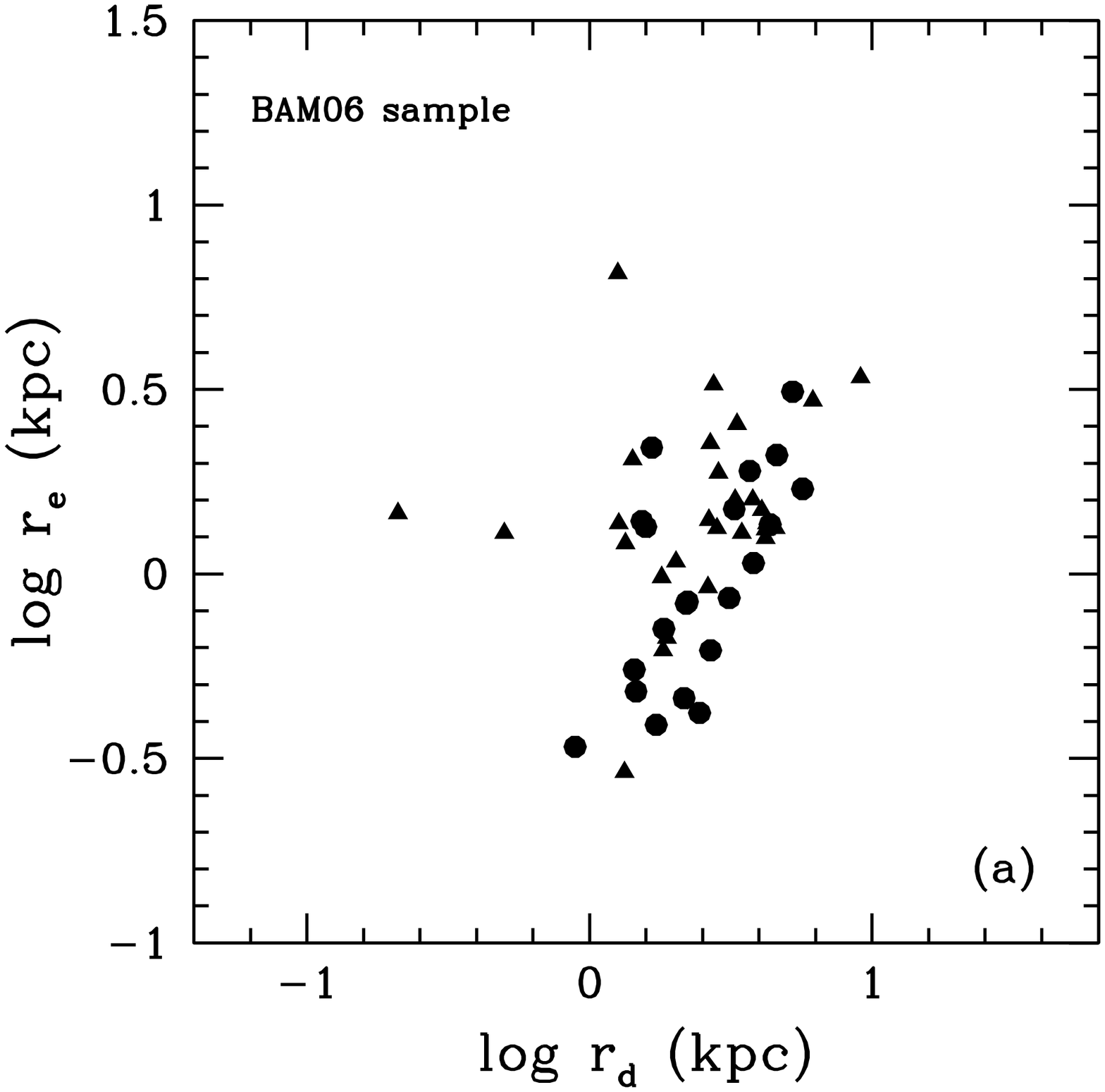}{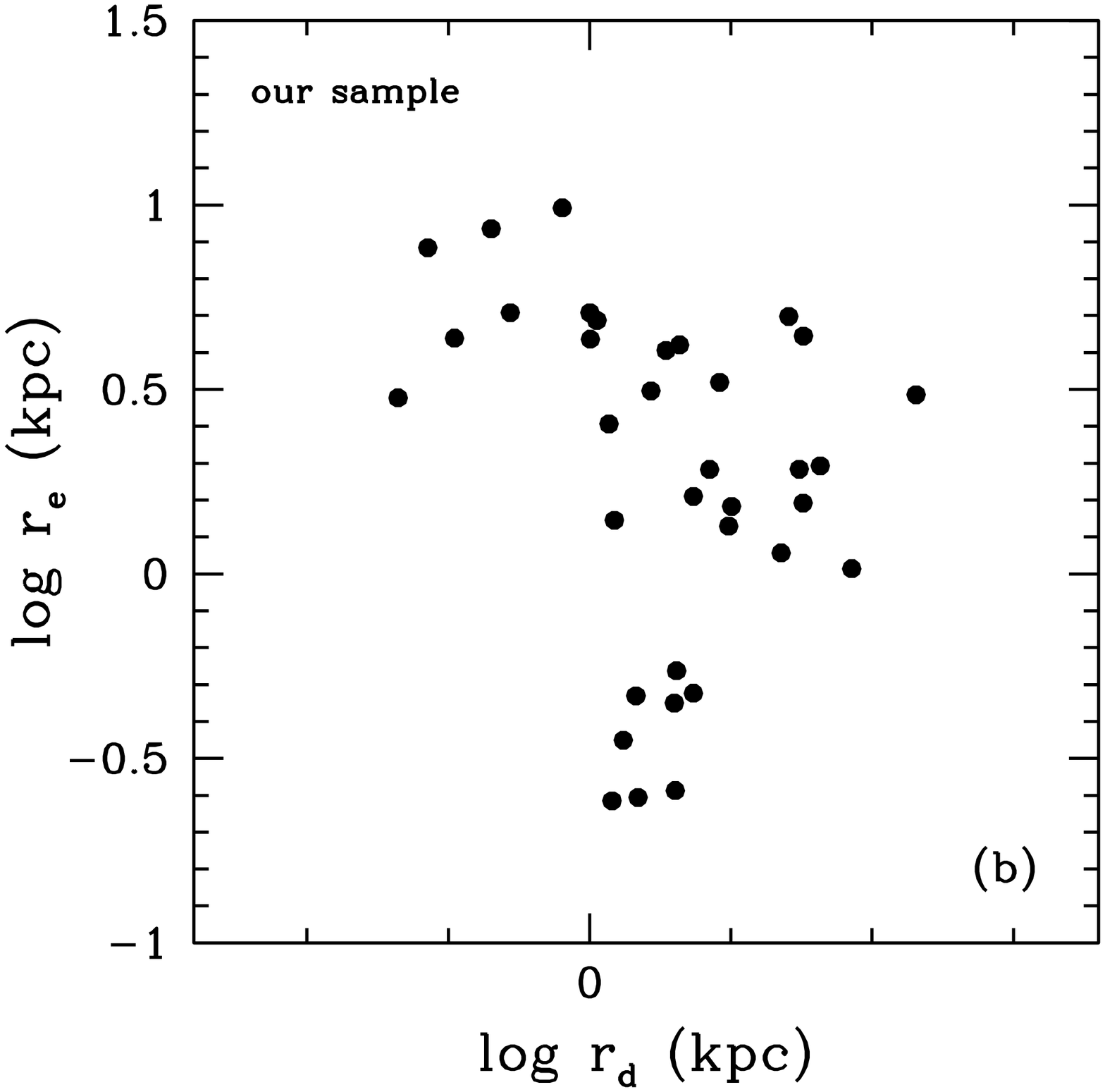}
\caption{The bulge effective radius ($\re$) plotted as a function of disk scale length ($\rd$) for lenticular galaxies (a) from BAM06 sample and (b) our sample. Lenticulars in the field and clusters are denoted by circles and triangles, respectively.}
\label{f2}
\end{figure}

In Figure~\ref{f2}(a)  we plot bulge effective  radius ($\re$) against
the disk scale length ($\rd$)  for the BAM06 sample, with open circles
representing  field  lenticulars  and  filled  triangles  representing
lenticulars  in  clusters. The  21  field  lenticulars  show a  strong
positive   correlation,  with linear correlation   coefficient  0.63   at  a
significance  level  better than  99.99\% \footnote{We have also used  
Spearman's  rank correlation coefficient to examine the trends reported 
in the the paper.  We find in every  case the significance 
of Spearman's coefficient is higher than the significance of the 
linear correlation coefficient we quote in the text.}. This  is  similar to  
the correlations reported for early-type  spiral 
galaxies  by Khosroshahi \etal \ (2000b) and for  late-type spiral galaxies 
by Courteau \etal \ (1996).  The 28 cluster  lenticulars have a correlation 
coefficient of 0.22  between $\re$  and $\rd$,  which is  significant at  
the 74.00\% level. While this level is too  low for the correlation to 
be accepted as  established, we  can say  that the  trend is  consistent  
with the significant  positive  correlation found  for  the field  lenticulars.
Moreover,  amongst  the cluster  lenticulars,  there  are three  clear
outliers.  Inspection of  their 2MASS K-band images shows  that one of
these (ESO  358-G59) has poor  signal-to-noise ratio, while  the other
two  (NGC 4638  and NGC  4787) are  obviously disk  dominated systems,
which are likely to have  disk scale lengths large than those reported
by BAM06.  We will omit  these three outliers from further discussion,
while noting  that our conclusions  are not changed by  this omission.
After  the omission,  the  $\re-\rd$ correlation  coefficient for  the
cluster lenticulars  increases to 0.61 with  significance greater than
99.99\%. 

We have shown in Figure~\ref{f2}(b)  the variation of $\re$ with $\rd$
for  our  sample  of  35  lenticulars.   A  majority (27/35)  of  field
lenticulars  from  the  sample  show  a  clear  anti-correlation  with
correlation coefficient  $-0.57$ at  a significance level  of 99.82\%.
All these lenticulars are found to have bulge luminosity exceeding the disk
luminosity,  with   mean  bulge-to-total  luminosity   ratio  $\langle
B/T\rangle = 0.63$. The  corresponding value for the BAM06 lenticulars
is $\langle B/T\rangle = 0.55$.  A separate group of lenticulars which
do not  follow the anti-correlation is  seen in the lower  part of the
diagram.  These are  found to be disk dominated  systems with low mean
bulge-to-total  luminosity  ratio $\langle  B/T\rangle  = 0.19$.   The
anti-correlation applies only to the bulge dominated systems.

\begin{figure}
\plotone{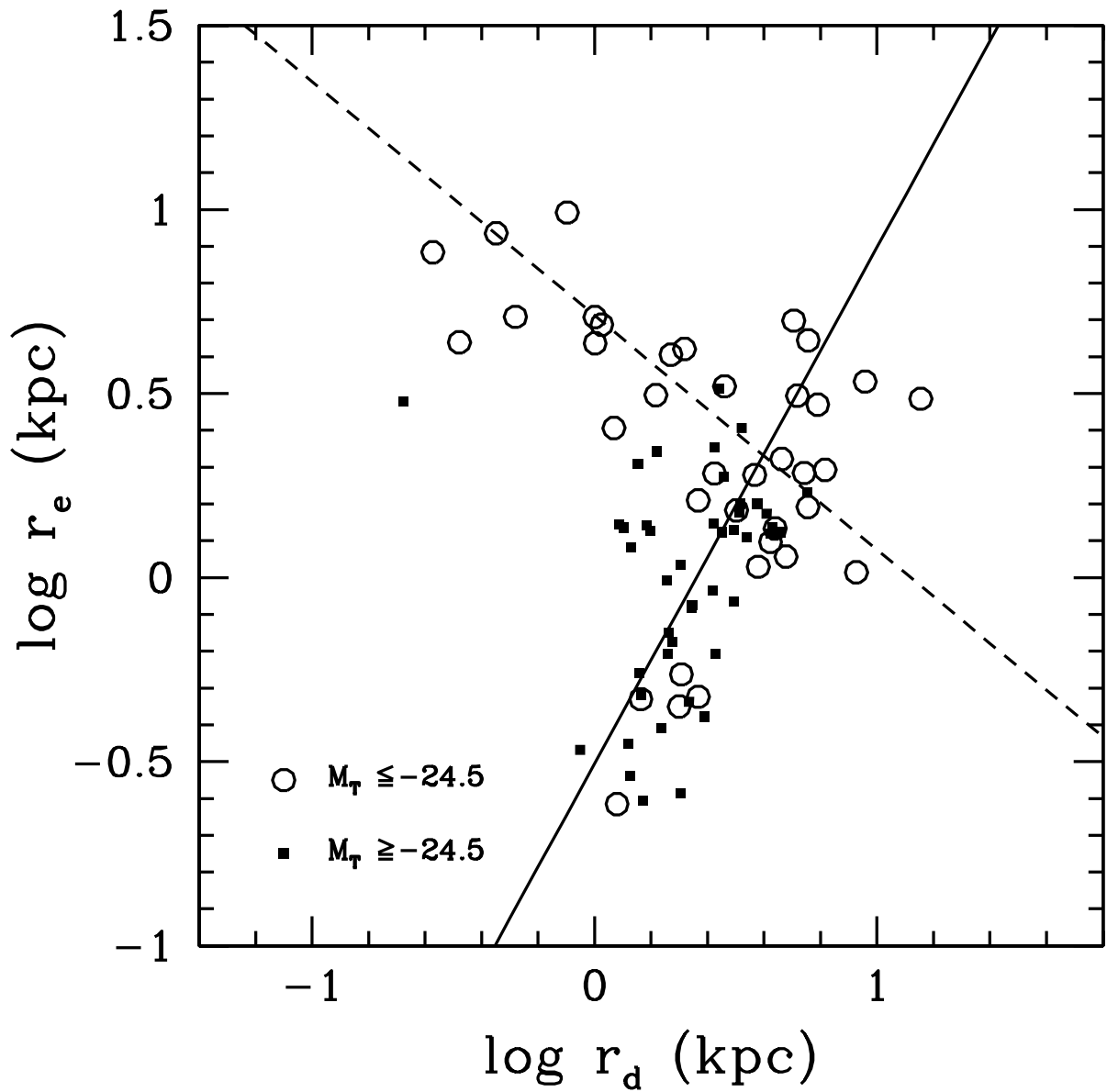}
\caption{Dependence of $\re$-$\rd$ relation on the total absolute magnitude in 
$K$ band for the combined sample (ours+BAM06) of lenticulars. Dashed line is the best-fit to the luminous lenticulars (open circles) excluding five outliers, with correlation coefficient $-0.64$ at 99.99\% significance level. Solid line is the best-fit to the less luminous lenticulars (filled squares, excluding one outlier) with correlation coefficient 0.48 at 99.89\% significance level.}
\label{f3}
\end{figure}

The difference between the BAM06 sample and our sample in the sign of
the $\re-\rd$ correlation is unlikely to be due to environmental
effects, since within the BAM06 sample, field lenticulars as well as
those in clusters show the same positive correlation.  However, as
galaxies in the BAM06 sample are systematically less luminous than
those in our sample, it could be that the fainter lenticular galaxies,
irrespective of their environment, show a positive $\re - \rd$
correlation, while the more luminous lenticulars exhibit a negative
correlation.  To test this hypothesis, we combine both samples and
then divide them into faint and bright groups, using $M_T=-24.5$ as
boundary.  The 43 faint lenticulars ($ M_T > -24.5$), with the outlier
indicated in Figure~\ref{f3} excluded, show positive correlation
between the bulge effective radius and disk scale length with
correlation coefficient 0.48 at a significance level better of
99.89\%.  On the other hand, the 32 bright lenticulars ($ M_T <
-24.5$), again excluding five obvious outliers, show a strong
anti-correlation with correlation coefficient $-0.64$ at a
significance level of $\> 99.99\%$ (see Figure~\ref{f3}).  We note
here that the boundary at $M_T=-24.5$ is merely indicative, and the
result does not critically depend on the choice of the dividing
luminosity.  Changing this value by half a magnitude on either side
leads to correlations significant at least at the $95\%$ level. We may
therefore say that the sense of the correlation is driven by the
luminosity of the galaxies, pointing to a possible fundamental
difference in the way in which faint and bright lenticulars are
formed.  The positive correlation seen in low luminosity galaxies
lends support to the hypothesis that such galaxies formed by the
stripping of gas from the halo and disk of late type spiral galaxies,
which formed their bulges through secular evolution.  More luminous
lenticulars likely formed their bulges differently, possibly through a
rapid collapse mechanism. We may mention here that dividing the entire
sample of lenticulars into bulge dominated and disk dominated systems,
using a boundary value of $B/T = 0.5$ or the median value 0.53 for the
whole sample, does not produce two well correlated subsamples. This is
consistent with the fact that a plot of $B/T$ against total luminosity
shows no obvious correlation.  The luminosity therefore seems to be
the main driver for the difference in the $\re-\rd$ correlation seen
in the population of lenticulars.

It has  been known  for some  time that the  {\it average}  \ S\'ersic
index $n$  systematically decreases from  early to late  type galaxies
(Andredakis, Peletier,  \& Balcells 1995; Khosroshahi  \etal \ 2000b).
This  decrease  is only  observed  in  a  statistical sense;  for  any
specific  Hubble type  there  is considerable  range  in the  measured
values  of $n$.   If  low luminosity  lenticulars  are indeed  evolved
late-type disk  galaxies, then  they should have  lower values  of the
S\'ersic index compared  to brighter lenticulars. We do  find that for
our combined sample, distribution of  the S\'ersic index for the faint
lenticulars  has  a  peak at  $n  \sim  3.25$  while that  for  bright
lenticulars has a peak at $n \sim 3.75$.

\section{Summary}

We find that lenticular galaxies show markedly different correlations 
between their bulge effective radius ($\re$)  and disk scale length 
($\rd$) as a function of their total luminosity. For faint lenticular 
galaxies ($ M_T > -24.5$), $\re$ and $\rd$ are positively correlated, in 
line with predictions of secular formation processes that likely formed 
the pseudo bulges of late-type disk galaxies. Such a formation scenario is 
also consistent with the predictions of numerical simulations of 
lenticular galaxy formation (Quilis, Moore, \& Bower 2000).  Bright 
lenticular galaxies with $ M_T < -24.5$, on the other hand, do not exhibit 
this correlation, indicating a different formation mechanism. These trends 
seem to hold irrespective of galaxy environment, although more luminous 
lenticulars are largely missing from our cluster sample. The relative 
fraction of lenticular galaxies is, of course, very different in clusters 
and in the field.

The correlations reported in this letter, need to be investigated with
larger  near-IR samples  of lenticular  galaxies, to  unravel possible
multiparametric  correlations   which  will  take   into  account  the
dependence  on  luminosity  and  environment and  provide  a  detailed
comparison with other galaxy types.   Such a study will be enabled in
the near future by new data such as the Large Area Survey component of
UKIDSS.

\acknowledgements

We thank  A.  Bedregal \etal \ for  providing us
their  bulge-disk  decomposition   results  in  electronic  form.  CDR
acknowledges the hospitality and  the Associateship provided by IUCAA.
We also thank the referee, A.  Arag{\'o}n-Salamanca, for insightful comments, 
which helped to improve 
the original manuscript.  YDM  thanks CONACyT (Mexico) for the project 
grant 39714-F, and IUCAA for hospitality provided during his visit.

\clearpage

\clearpage

\clearpage

\end{document}